\documentclass[prx,twocolumn,superscriptaddress,citeautoscript,showpacs,amsart]{revtex4-2}

\usepackage{blindtext}
\usepackage{graphicx}
\usepackage{bm}
\usepackage{epsfig}
\usepackage{amssymb}
\usepackage{amsfonts}
\usepackage{braket}
\usepackage{color}
\usepackage{epstopdf}
\usepackage{hyperref}
\usepackage{float}
\restylefloat{table}
\usepackage{bibentry}
\usepackage{color}
\usepackage{multirow}
\usepackage[caption=false]{subfig}

\usepackage{amsfonts}
\usepackage{amsthm}
\usepackage{graphicx}
\usepackage{multirow}
\usepackage{color}
\usepackage{bbold}
\usepackage{bm}
\usepackage{times}
\usepackage{amsmath,bm,amsfonts}
\usepackage{dcolumn}
\usepackage{graphicx}
\usepackage{latexsym}

\newcommand{\bpm}{\begin{pmatrix}}
\newcommand{\epm}{\end{pmatrix}}
\newcommand{\ba}{\begin{eqnarray}}
\newcommand{\ea}{\end{eqnarray}}
\newcommand{\bd}{\begin{displaymath}}

\graphicspath{{figures/}}

\begin{document}
\title{Evolution of electronic band reconstruction in thickness-controlled perovskite SrRuO$_3$ thin films}

\author{Byungmin Sohn}
\affiliation{Center for Correlated Electron Systems, Institute for Basic Science, Seoul 08826, Korea}
\affiliation{Department of Physics and Astronomy, Seoul National University, Seoul 08826, Korea}

\author{Changyoung Kim}
\email[Electronic address:$~~$]{changyoung@snu.ac.kr}
\affiliation{Center for Correlated Electron Systems, Institute for Basic Science, Seoul 08826, Korea}
\affiliation{Department of Physics and Astronomy, Seoul National University, Seoul 08826, Korea}

\date{\today}

\begin{abstract}

Transition metal perovskite oxides display a variety of emergent phenomena which are tunable by tailoring the oxygen octahedral rotation. SrRuO$_3$, a ferromagnetic perovskite oxide, is well-known to have various atomic structures and octahedral rotations when grown as thin films. However, how the electronic structure changes with the film thickness has been hardly studied. Here, by using angle-resolved photoemission spectroscopy and electron diffraction techniques, we study the electronic structure of SrRuO$_3$ thin films as a function of the film thickness. Different reconstructed electronic structures and spectral weights are observed for films with various thicknesses. We suggest that octahedral rotations on the surface can be qualitatively estimated via comparison of intensities of different bands. Our observation and methodology shed light on how structural variation and transition may be understood in terms of photoemission spectroscopy data.

\end{abstract}
\maketitle

\section {introduction}
Transition metal oxides (TMO) have been intensively investigated over the past decades owing to their novel phenomena such as high-temperature superconductivity, multiferroicity, Mott transition, and magnetism~\cite{tokura2000orbital, damascelli2003angle, spaldin2019advances, ahn2021designing}. These phenomena are often understood to emerge from the strongly correlated $d$ electrons in which the fundamental degrees of freedom -- spin, orbital, charge, and lattice -- are entangled. Especially, in Ruddlesden-popper (RP) oxide perovskites, which have a chemical formula of A$_{n+1}$M$_{n}$O$_{3n+1}$ ($n$~=~1,~2,~3~$\cdots$), transition metal $d$ electrons are combined with oxygen $p$ electrons, forming MO$_6$ oxygen octahedron electronic structure. Controlling the structure of oxygen octahedra determines the metallicity and magnetism in many RP oxides such as Sr$_2$RuO$_4$~\cite{anisimov2002orbital}, Sr$_2$IrO$_4$~\cite{kim2009phase}, and SrRuO$_3$ (SRO)~\cite{koster2012structure, kim2022heteroepitaxial}.

In particular, MO$_6$ octahedron rotation significantly alters the electronic structure and may induce phase transitions, often accompanying changes in physical properties. A prime example is a well-known ferromagnetic metallic TMO perovskite SRO for which various characteristics appear with RuO$_6$ octahedral rotations; a large octahedral rotation angle favors an insulating antiferromagnetic phase whereas a small octahedral rotation prefers a metallic ground state with ferromagnetism~\cite{ahn1999spectral, jin2008high,ali2022origin}. As such, numerous attempts to control the MO$_6$ octahedron angle and characteristics of TMO systems have been reported~\cite{jin2008high,vailionis2008room,lu2015strain,balachandran2013interplay}. All these show that MO$_6$ octahedral rotation can play an important role in modifying the intrinsic properties. Accordingly, finding suitable experimental methods to control oxygen octahedra in TMOs has been of great interest. 

One of the well-adopted experimental methods for tuning the MO$_6$ octahedra is growing TMOs in thin films. Due to the strain and interfacial effects induced by the substrate, thin films may have octahedral rotation different from that of the bulk~\cite{hwang2012emergent}. Since thin films tend to have similar characteristics with bulk when the thin films are thick enough, variation in the octahedral rotation (and, crystal symmetry) as a function of the thickness of thin films has been studied~\cite{siwakoti2021abrupt,chang2011thickness, roh2021structural}. However, most studies were focused on atomic structures through X-ray diffraction analysis, whereas only a few studies of thickness-dependent electronic structures have been reported so far.

\begin{figure*}[]
	\includegraphics[width=\linewidth]{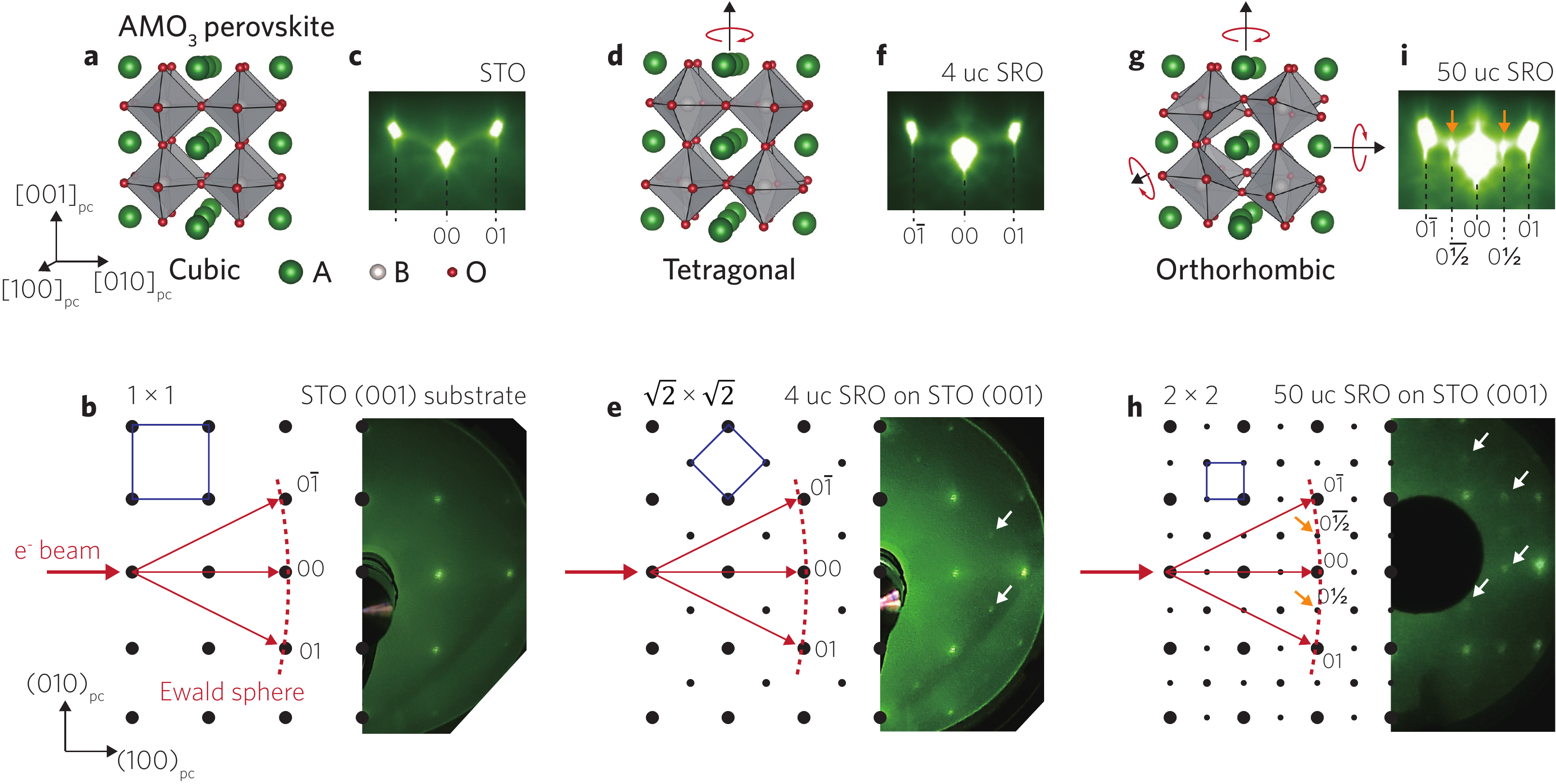}
	\centering
	\caption{{\bf Reflection high-energy electron diffraction (RHEED) and low-energy electron diffraction (LEED) patterns of various atomic structures.} 
		Schematics of ({\bf a}) cubic, ({\bf d}) tetragonal, and ({\bf g}) orthorhombic structures of perovskites. RHEED patterns of ({\bf c}) SrTiO$_3$ (STO) (001) substrate, ({\bf f}) 4~unit-cell (uc), and ({\bf i}) 50~uc SrRuO$_3$ (SRO) thin films on grown on STO substrates are given. RHEED patterns are measured along the (100)$_{\rm pc}$ crystalline direction. [0~$\frac{1}{2}$] peaks marked with orange arrows appear due to the 2~$\times$~2 reconstruction. 
		LEED schematics and patterns of ({\bf b}) STO (001) substrate, ({\bf e}) 4~uc, and ({\bf h}) 50~uc SRO thin films on STO substrates are plotted. LEED measurements were done with an electron energy of 150~eV at 300~K. The 4~uc and 50~uc SROs show  $\sqrt2$~$\times$~$\sqrt2$ and 2~$\times$2 reconstruction, respectively (marked with white arrows). When the electron ($e^-$) beam comes along the (100)$_{\rm pc}$ direction, [0~$\frac{1}{2}$] peaks are observed in the RHEED pattern of orthorhombic perovskite structure ({\bf i}) due to the 2~$\times$~2 reconstruction. {\bf f}, {\bf i}, and {\bf h} are adapted from ref.~\cite{sohn2021sign}.}
	\label{fig:1}
\end{figure*}
Here, we investigate atomic and electronic structures of SRO thin films grown on SrTiO$_3$ (STO) (001) substrates as a function of the thickness with {\it in-situ} reflection high-energy electron diffraction (RHEED), low-energy electron diffraction (LEED) and angle-resolved photoemission spectroscopy (ARPES). Atomic structural transitions appear with thickness variation, and reconstructed electronic structures are observed in ARPES data. Based on our simple theoretical model, we claim that the intensity of the reconstructed band structure is intimately related to oxygen octahedral rotation.\\

\section {Methods}

SRO thin films with thicknesses of 2.5, 4, 5, 10, 30, and 50~unit-cell (uc) were grown on TiO$_2$-terminated STO (001) substrates with RHEED-monitored pulsed laser deposition (PLD). Here, a 2.5~uc film consists of 50 $\%$ of 2-uc-thick SRO and 50 $\%$ of 3-uc-thick SRO. Since SRO is grown by step-flow mode, the fractional thickness can be intentionally obtained by stopping the deposition in the middle of the growth process~\cite{sohn2021sign,wang2020controllable}. Before the growth, STO substrates were prepared with deionized water etching and then {\it in-situ} pre-annealing at 1,070~$^\circ$C for 30 minutes under the oxygen partial pressure of 5~$\times$~10$^{-6}$~torr. SRO thin films were grown under the oxygen partial pressure of 100~mTorr at 700~$^\circ$C. A KrF excimer laser (248~nm) irradiated an SRO target with a fluence of 1-2~J/cm$^2$ and a repetition rate of 2~Hz. Samples were {\it in-situ} transferred for ARPES and LEED measurements. ARPES measurements were performed at 10~K equipped with a hemispherical electron analyzer and a He discharge lamp. He-I$\alpha$ light with the photon energy of 21.2~eV was used. Before ARPES measurements, samples were annealed at 550~$^\circ$C to remove surface contaminants. \\

\section {Results}

We measured RHEED and LEED of a bare STO (001) substrate, 4~uc and 50~uc SRO on STO (001) substrates at room temperature. Here, electrons come through the (001)$_{\rm pc}$ and (100)$_{\rm pc}$ direction for LEED and RHEED, respectively. Since STO has a cubic structure at room temperature (with a symmetry group of {\it Pm-3m} and a glazer notation of a$^0$a$^0$a$^0$)~\cite{rimai1962electron}, oxygen octahedra are not expected to be rotated in an STO substrate at all (Fig. 1a). Thus, the LEED pattern of STO substrates shows a square-shaped Brillouin zone without any reconstruction (Fig. 1b). Bright three spots from 0$\overline{1}$, 00, and 01 peaks, crossing an Ewald sphere, are observed in the RHEED pattern (Fig. 1c).

On the other hand, the 4~uc SRO on an STO substrate is expected to have a tetragonal ({\it I4/mcm}, a$^0$a$^0$c$-$) structure with in-plane octahedral rotation (Fig. 1d)~\cite{sohn2021stable}, resulting in in-plane unit-cell doubling with the $\sqrt{2}$~$\times$$\sqrt{2}$ periodicity. $\sqrt{2}$~$\times$$\sqrt{2}$ reconstruction spots appear in LEED patterns (white arrows in Fig. 1e). Since an Ewald sphere is not crossing the $\sqrt{2}$~$\times$$\sqrt{2}$ spots with our RHEED geometry, the same RHEED patterns are observed for the bare STO and 4~uc SRO (Fig. 1f).

It has been reported that a thickness-dependent structural phase transition (from tetragonal to orthorhombic phases) occurs in SRO thin films on an STO substrate when SRO films become thicker than 19~uc~\cite{chang2011thickness}. Thus, the 50~uc SRO film is expected to be orthorhombic ({\it Pbnm}, a$^-$b$^+$a$^-$) with in-plane and out-of-plane octahedral rotation (Fig. 1g), accompanied by in-plane unit-cell doubling with 2~$\times$~2 reconstruction. Our LEED pattern clearly shows 2~$\times$~2 spots (orange and white arrows in Fig. 1h). Contrary to the tetragonal case, an Ewald sphere is crossing the 2~$\times$~2 reconstruction spots, which are observed in the RHEED pattern (marked with orange arrows in Fig. 1i).
 

\begin{figure*}[]
\includegraphics[width=0.90\linewidth]{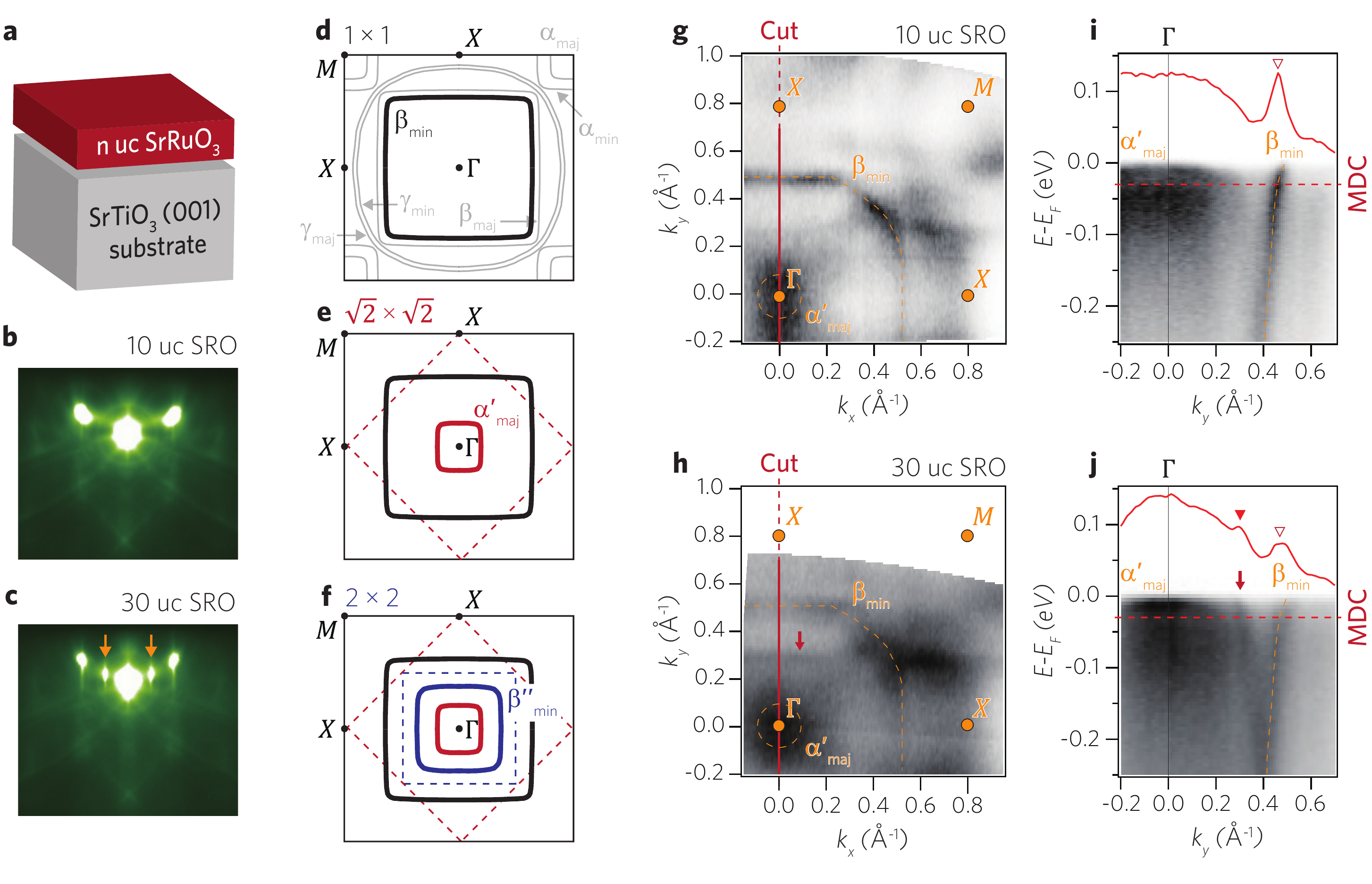}
\centering
\caption{{\bf Observation of $\sqrt2$~$\times$$\sqrt2$ and 2~$\times$~2 reconstructed bands in SRO thin films.} 
         {\bf a} A schematic of $n$~uc SRO thin films on an STO (001) substrate. {\bf b,c} RHEED patterns of 10 and 30~uc SRO thin films on STO substrates. [0~$\frac{1}{2}$] peaks are marked with orange arrows. 
         {\bf d-f} Fermi surface (FS) schematics of cubic (1~$\times$~1, {\bf d}), tetragonal ($\sqrt2$~$\times$$\sqrt2$, {\bf e}), and orthorhombic (2~$\times$~2, {\bf f}) SRO thin films. In {\bf d}, $\beta_{\rm min}$ is represented with black solid lines, whereas other bands are plotted with gray solid lines. In {\bf e} and {\bf f}, $\alpha^{'}_{\rm maj}$ and $\beta^{''}_{\rm min}$ bands are represented with red and blue solid lines, respectively. To make it easy to understand, we only plot specific bands in {\bf e} and {\bf f}.
         {\bf g,h} FSs of 10 and 30~uc SRO thin films. 
         {\bf i,j} Energy-momentum cuts of 10 and 30~uc SRO thin films along the $\Gamma$-X direction ($k_x$~=~0). Momentum distribution curves (MDCs) at $E$~=~$E_F$~-~30~meV, where $E_F$ is the Fermi level, are also plotted (red curves). The open (closed) inverted triangle represents a peak from $\beta_{\rm min}$ ($\beta^{''}_{\rm min}$) band. Red arrows in {\bf h} and {\bf j} indicate $\beta^{''}_{\rm min}$ band.
}
        
\label{fig:2}
\end{figure*}

We next compare electronic structures of tetragonal and orthorhombic SRO thin films by performing ARPES on 10~uc and 30~uc SRO thin films (Fig. 2a). To investigate the structural phases of each film, RHEED patterns were observed after the growth process (Fig. 2b and 2c). The 30~uc SRO thin film shows 2~$\times$~2 reconstruction represented by orange arrows in Fig. 2c, which supports that 10~uc (30~uc) SRO is tetragonal (orthorhombic) consistent with a previous report~\cite{chang2011thickness}.

Because Ru 4$d$ states cross the Fermi level ($E_F$), the electronic structure of SRO is often described within the $t_{2g}$ manifold with four electrons~\cite{chen2013weyl}. The $t_{2g}$ manifold consists of six states ($d_{xz}$, $d_{yz}$, $d_{xy}$ with majority (maj) and minority (min) spins) due to the ferromagnetism and six bands are expected to cross $E_F$. Based on previous reports~\cite{shai2013quasiparticle,sohn2021sign,hahn2021observation}, we present FS schematics as depicted in Figs. 2d-f. Without octahedral rotation (Fig. 2d), two hole-like spin-resolved bands ($\alpha_{\rm maj}$, $\alpha_{\rm min}$) and four electron-like spin-resolved bands ($\beta_{\rm maj}$, $\beta_{\rm min}$, $\gamma_{\rm maj}$, $\gamma_{\rm min}$) are expected to cross $E_F$. In tetragonal SRO films, in-plane octahedral rotation induces $\sqrt{2}$~$\times$$\sqrt{2}$ reconstruction and the Brillouin zone is reduced to the one indicated by a red-dotted square (Fig. 2e). When SRO films become orthorhombic and have both in-plane and out-of-plane octahedral rotations, the Brillouin zone is further reduced to the one shown by a blue-dotted square (Fig. 2f) due to the 2~$\times$~2 reconstruction. 

Figures 2g and 2h show measured Fermi surfaces (FSs) of 10~uc and 30~uc SRO films. Note that due to matrix element effects only part of the band structure is often visible in ARPES measurements~\cite{moser2017experimentalist}. We noticed that the $\beta_{\rm min}$ band is clearly observed in both FS maps (corresponds to the black pocket in Fig. 2d). Note that the Fermi momentum ($k_F$) of the $\beta_{\rm min}$ band is almost the same ($k_F$~=~0.51~${\rm \AA^{-1}}$ along $k_y$~=~0) in the two maps. Strong spectral weight is observed at the $\Gamma$ point which originates from the $\alpha'_{\rm maj}$ band observed due to $\sqrt{2}$~$\times$$\sqrt{2}$ reconstruction (the red pocket in Fig. 2e)~\cite{sohn2021sign}. In Fig. 2h, interestingly, additional spectral feature (marked with a red arrow) appears, which is not observed for the 10~uc SRO FS.

To examine the additional spectra observed in the 30~uc SRO FS, energy-momentum high-symmetry cuts along $k_x$~=~0 (red solid lines in Figs. 2g and 2h) are presented in Figs. 2i and 2j. Strong spectra from $\alpha'_{\rm maj}$ and the $\beta_{\rm min}$ band are clearly observed in both high-symmetry cuts, whereas an additional dispersive band (marked with a red arrow) is observed in the 30~uc SRO. Momentum distribution curve (MDC) at $E$~=~$E_F$~-~30~meV is plotted for each figure, showing an additional peak (marked with a closed inverted triangle) appears in 30~uc SRO. As the additional dispersive band and $\beta_{\rm min}$ band are symmetric with respect to $k_y$~=~0.4 ${\rm \AA^{-1}}$, we conclude the additional dispersive band is $\beta''_{\rm min}$ band, the blue pocket in Fig. 2f, induced by 2~$\times$~2 reconstruction.

\begin{figure*}[]
\includegraphics[width=0.94\linewidth]{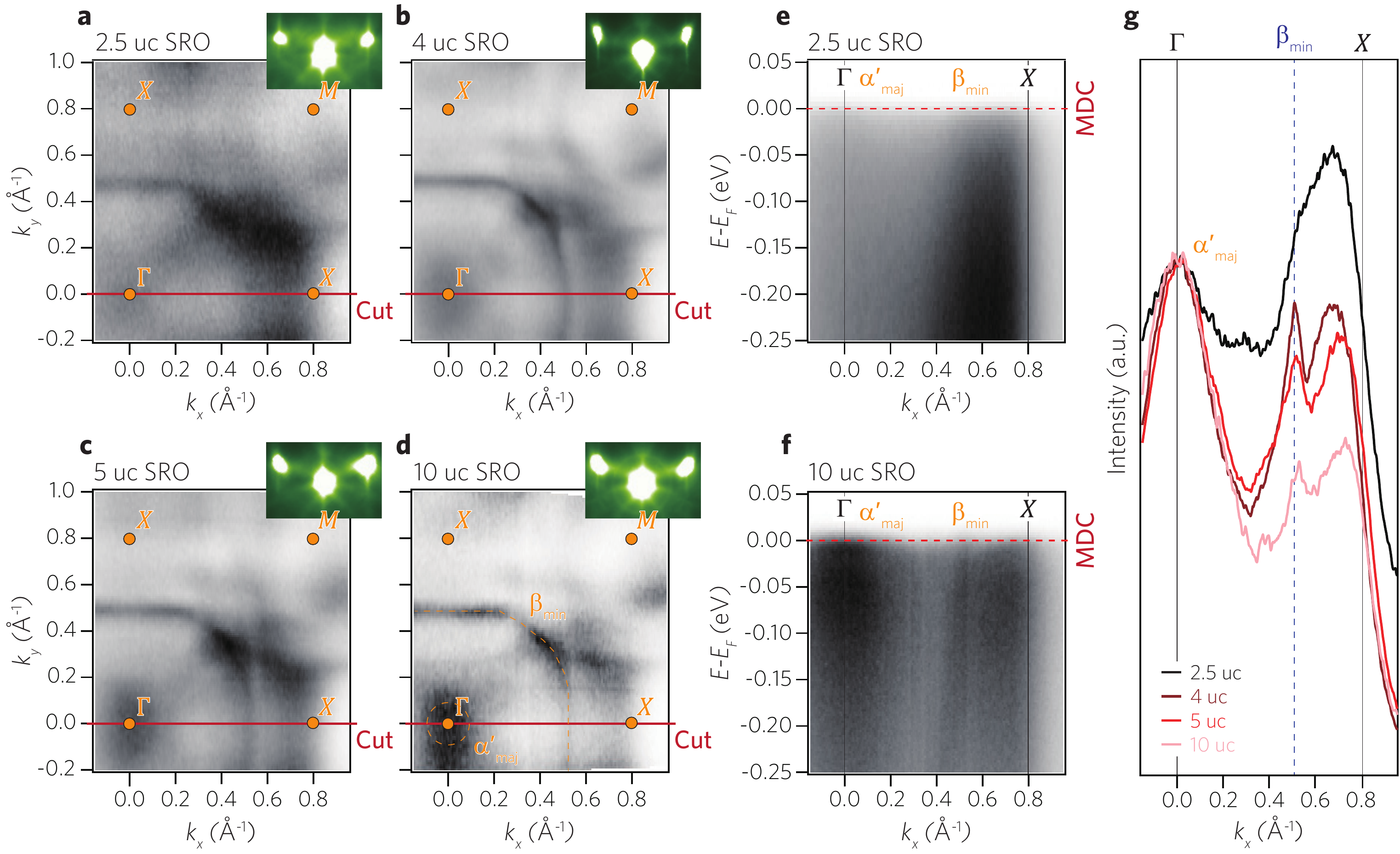}
\centering
\caption{{\bf Observation of thickness-dependent intensity of $\sqrt2$~$\times$$\sqrt2$ reconstructed bands.} 
         {\bf a-d} FSs of $n$~uc ($n$~=~2.5, 4, 5, and 10) SRO thin films grown on STO (001) substrates. (Inset) RHEED patterns measured along the (100)$_{pc}$ direction.         
         {\bf e,f} Energy momentum cuts of 2.5 and 10~uc SRO thin films along the $\Gamma$-X direction ($k_y$~=~0).
         {\bf g} MDCs along $k_y$~=~0 at $E$~=~$E_F$ as a function of the thickness. Peaks from $\beta_{\rm min}$ band are marked with a blue dotted line. All MDCs are normalized with the intensity at the $\Gamma$ point.
		{\bf b} is adapted from ref.~\cite{sohn2021sign}.
     	 }
\label{fig:3}
\end{figure*}

We also investigate how the electronic structures vary as a function of the thickness for SRO films within the tetragonal structure. Figures 3a-d show FSs of 2.5, 4, 5, and 10~uc SRO films. Based on the RHEED patterns in inset figures and results in previous reports~\cite{sohn2021observation,chang2011thickness,sohn2021stable}, we determine that all of these SRO films are in tetragonal structure. We note that, regardless of the thickness, the fermiology of the films remains the same. Figures 3e and 3f show energy-momentum band dispersion of 2.5 and 10~uc SRO films along $k_y$~=~0. A spectral weight near the $\Gamma$ point is strongly observed in the 10~uc SRO film, whereas the spectral weight is suppressed in the 2.5~uc SRO film.  

We plot MDCs of these tetragonal SRO films along $k_y$~=~0 at $E_F$ (Fig. 3g). Three peaks are seen in each MDC. As we discussed above for Fig. 2, the spectral weight at the $\Gamma$ point ($k_x$~=~0.51~${\rm \AA^{-1}}$) stems from $\alpha'_{\rm maj}$ ($\beta_{\rm min}$). Notably, the peak ratio between $\alpha'_{\rm maj}$ and $\beta_{\rm min}$ becomes larger as the thickness increases. Meanwhile, the position of the $\beta_{\rm min}$ peak remains almost the same at $k_x$~=~0.51~${\rm \AA^{-1}}$, regardless of the thickness.\\

\section {Discussion}

Now, we wish to discuss our observations. ARPES results from 2.5, 4, 5, and 10~uc tetragonal SRO films do not show significant variation in the band structure but the spectral weight intensity differs. Since electronic structures are highly sensitive to the structure of MO$_6$ octahedra in SRO systems~\cite{kim2022signature, ko2007strong}, FS should appear different if MO$_6$ octahedral rotations are different. In fact, ARPES is highly surface-sensitive; less than 1 nm ($\sim$~2.5~uc of SRO layers) is expected to be probed with 21.2~eV photon~\cite{seah1979quantitative}. Thus, our observations indicate that the atomic structure near the surface may remian almost the same regardless of the thickness, resulting in the same fermiology and $k_F$. 

However, if the atomic structure is exactly the same, the different intensity ratios in MDCs (Fig. 3g) cannot be explained. Since SRO films we studied are ultrathin ones and the lattice constant mismatch between bulk STO and SRO is 0.06~\%~\cite{koster2012structure}, SRO films are expected to be fully strained, {\it i.e.} the in-plane lattice constant of the SRO films is the same as that of STO substrate~\cite{sohn2021stable,chang2011thickness,vailionis2008room}. Instead, we speculate that the MO$_6$ octahedral rotation angle near the surface layers is different. Namely, we claim that the octahedral rotations are different among the SRO films we studied. The difference in the octahedral rotation angle must be responsible for the difference in the intensity ratio.

\begin{figure}[]
	\includegraphics[width=0.95\linewidth]{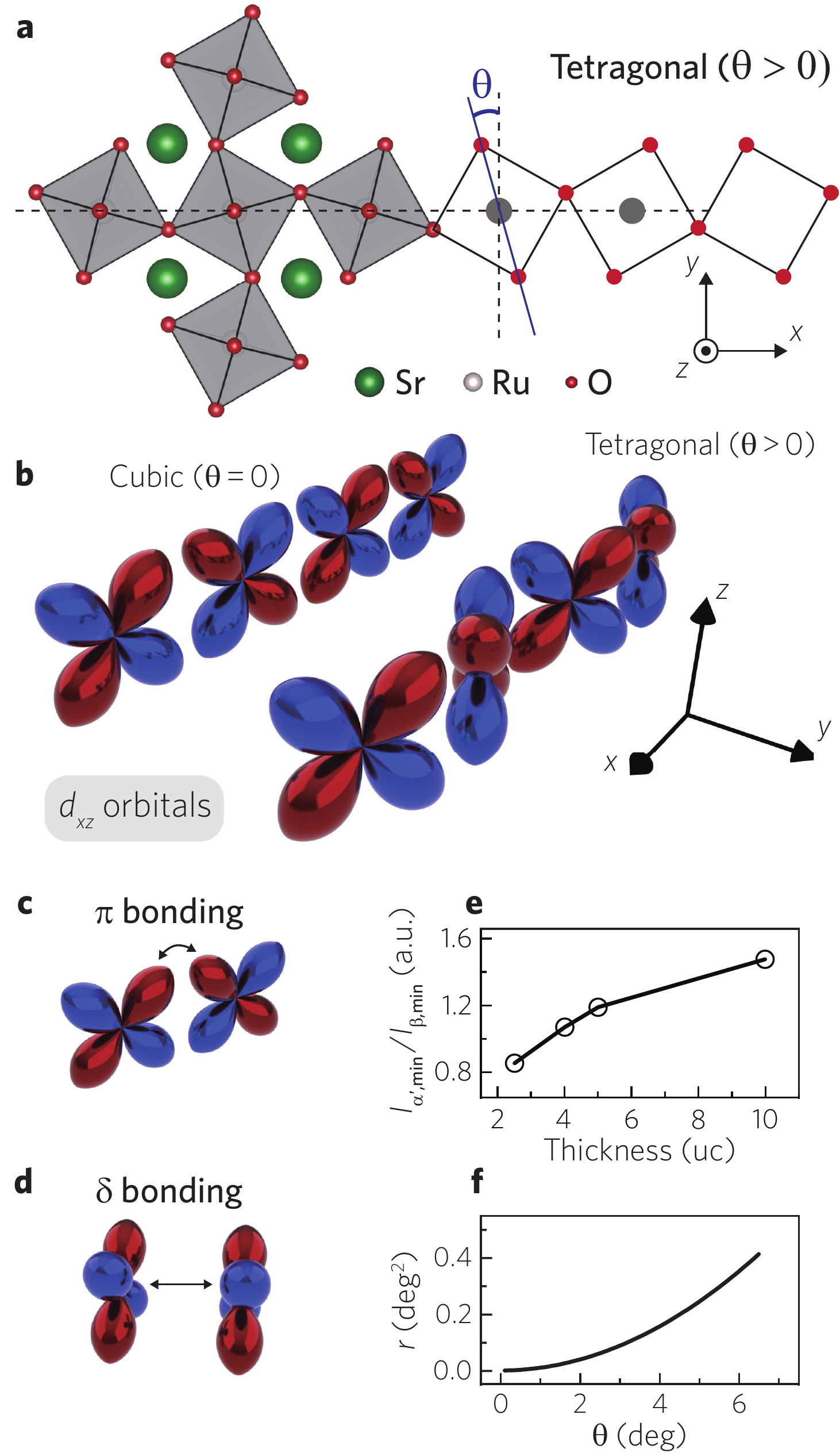}
	\centering
	\caption{{\bf Intensity variation of reconstructed bands with MO$_6$ octahedral rotation.} 
		{\bf a} Schematic of a single tetragonal SRO layer with an in-plane octahedral rotation angle, $\theta$. 
		{\bf b} Schematic of $d_{xz}$ orbitals with and without in-plane oxygen octahedral rotation.
		{\bf c,d} Schematics of $\pi$- and $\delta$-bondings between two $d_{xz}$ orbitals.
		{\bf e} Intensity ratio between the $\sqrt2$~$\times$$\sqrt2$ reconstructed and original band as a function of the thickness. $I_{\alpha',{\rm maj}}$ and $I_{\beta,{\rm min}}$ are defined as $E_F$ MDC intensity at the $\Gamma$ point and the peak position of $\beta_{\rm min}$ band, respectively.         
		{\bf f} $r$ vs $\theta$ plot where $r$ is defined as $\theta^2$($t_{\delta}$/$t_{\pi}$)$^2$. $t_{\delta}$ and $t_{\pi}$ are hopping integrals of $\delta$- and $\pi$-bonding electrons between two $d_{xz}$ orbitals.}
	\label{fig:4}
\end{figure}

We attempt to explain the thickness-dependent suppression of the spectral weight near the $\Gamma$ point with a simple toy model. Especially, we discuss the intensity suppression along $k_x$~=~0 given in Fig. 3g. The spectral weight near the $\Gamma$ point stems from $\alpha'_{\rm maj}$; thus, $d_{xz}$ and $d_{yz}$ orbitals mainly contribute to the spectral weight~\cite{sohn2021sign}. Since the electron hopping integral between $d_{yz}$ orbitals along the x-axis is approximately ten times smaller than that of between $d_{xz}$ orbitals~\cite{mazin2000tight}, we only consider $d_{xz}$ orbitals in our model. Figure 4a is a schematic of a single in-plane tetragonal SRO layer. We define $\theta$ as the in-plane octahedral rotation angle. Chains of $d_{xz}$ orbitals in an SRO layer are shown in Fig. 4b for cubic ($\theta$~=~0) and tetragonal ($\theta$~\textgreater~0) structure. When $\theta$ is zero, $d_{xz}$ orbitals are not rotated, so electrons in $d_{xz}$ hop only through a $\pi$ bonding (Fig. 4c). On the other hand, $d_{xz}$ orbitals rotate when $\theta$~\textgreater~0 and electrons hop through both $\pi$ and $\delta$ bondings (Fig. 4d). When the $d_{xz}$ chain is equally spaced but $d_{xz}$ orbitals are distinct and rotated, the intensity ratio between the folded and original band (or, the low and high intensity band) along the $x$-axis is expressed as 

\begin{equation}
	\frac{I'}{I} =  \frac{(\braket{ k_f|d_{xz,\theta}}-\braket{ k_f|d_{xz,-\theta}})^2}{(\braket{ k_f|d_{xz,\theta}}+\braket{ k_f|d_{xz,-\theta}})^2},
\end{equation}
where $I'$, $I$, and $\ket{k_f}$ denote the intensity of a folded band, the intensity of the original band, and a single plane wave component with a momentum $k_f$, respectively~\cite{moser2017experimentalist}. $\ket{d_{xz,\theta}}$ and $\ket{d_{xz,-\theta}}$ are a $d_{xz}$ orbital rotated by $\theta$ in the clockwise and counter-clockwise direction, respectively. With an assumption that $\theta$ is small, $\braket{ k_f|d_{xz,\pm \theta}} \simeq \braket{ k_f|d_{xz} \pm \theta d_{yz}}$, where $d_{xz}$ ($d_{yz}$) represents $d_{xz,\pm \theta}$ ($d_{yz,\pm \theta}$) with $\theta$~=~0. Thus, the intensity ratio between the bands with the $d_{xz}$ orbital component is expressed as

\begin{equation}
	\frac{I'}{I} \simeq (\frac{\braket{ k_f|\theta d_{yz}}}{\braket{k_f|d_{xz}}})^2 \propto \theta^2(\frac{t_\delta}{t_\pi})^2,
\end{equation}
where $t_\pi$ ($t_\delta$) denotes an electron hopping integral between the same $d_{xz}$ orbitals through $\pi$ ($\delta$) bonding. Note that along the $y$-axis, the electron hopping integral between $d_{xz}$ orbitals is approximately ten times smaller than that of between $d_{yz}$ orbitals~\cite{mazin2000tight}. Thus, the intensity suppression can be well explained with chains of $d_{yz}$ orbitals instead of those of $d_{xz}$ orbitals. For the y-axis case, the $d_{yz}$ ($d_{xz}$) orbital is substituted by $d_{xz}$ ($d_{yz}$) in equations.

We compare our experimental observation with the theoretical equation. The intensity ratio between the $\alpha'_{\rm maj}$ and $\beta_{\rm min}$ band spectral weights, $I_{\alpha',{\rm maj}}$/$I_{\beta,{\rm min}}$, is plotted in Fig. 4e, and $r$~=~$(\theta t_\delta / t_\pi)^2$ is shown as a function of $\theta$ in Fig. 4f. We assume a value of 0.1 for $t_\delta / t_\pi$ based on the reported Slater-Koster parameters in bulk SRO~\cite{mazin2000tight}. As the SRO thin film becomes thicker, $I_{\alpha',{\rm maj}}$/$I_{\beta,{\rm min}}$ increases. Similarly, $r$ also increases with $\theta$. Since $I_{\alpha',{\rm maj}}$/$I_{\beta,{\rm min}}$ is proportional to $r$, we claim that the large (small) intensity ratio stems from the large (small) octahedral rotation angle.

It has been reported that octahedral rotations in ultrathin films highly depend on the thickness due to the interfacial effects; the MO$_6$ octahedra of films tend to adopt the same octahedral rotations of the substrate~\cite{fowlie2019thickness,liao2016controlled}, which relaxes as it propagates through the film. Since STO has a smaller octahedral rotation angle ($\theta$~$\sim$~2~deg at 10~K~\cite{unoki1967electron,muller1968characteristic,tsuda1995refinement}) than that of bulk SRO ($\theta$~$\sim$~6~deg at room temperature~\cite{gao2016interfacial}), we expect that thicker SRO films have a larger rotation angle on the surface than thinner films. We suggest that our observation of the thickness-dependent intensity ratio variation can be an indication of different octahedral rotation on the surface of SRO ultrathin films. Further studies may be desired to address the issue in a more quantitative manner.\\

\section {Conclusion}
In summary, we investigated how the electronic structure on the surface of SRO thin film changes with the atomic reconstruction through thin-film thickness dependent studies. Differently reconstructed band structures were observed in orthorhombic (thick) and tetragonal (thin) SRO films. As for SRO ultrathin films (n~$\leqq$~10~uc), they show almost thickness independent Fermi surface topology but exhibit different spectral weight intensities. Based on our theoretical model, we conclude that the rotation angle of the surface octahedra becomes smaller as the thickness is reduced in the ultrathin limit. Our observation and methodology may shed light on how structural change and transition can be understood in the view of photoemission spectroscopy and electronic structures.

\section*{Acknowledgements}
We gratefully acknowledge insightful discussions with Younsik Kim. This work is supported by the Institute for Basic science in Korea (Grant No. IBS-R009-G2). CK acknowledges the supported by the National Research Foundation of Korea(NRF) grant funded by the Korea government(MSIT). (No. 2022R1A3B1077234)\\


%

\end{document}